\documentclass[%
reprint,
superscriptaddress,
nofootinbib,
 amsmath,amssymb,
 aps,
]{revtex4-2}

\usepackage{graphicx}
\usepackage{dcolumn}
\usepackage{bm}
\usepackage{hyperref}
\hypersetup{
    colorlinks = true,
    citecolor = blue,
}
\usepackage[mathlines]{lineno}
\usepackage[normalem]{ulem}
\usepackage{cancel}

\usepackage{lipsum}
\usepackage{tensor}
\usepackage[capitalise]{cleveref}
\usepackage{mathtools}

\crefname{section}{Sec.}{Sec.}

\begin{document}

\preprint{APS/123-QED}

\title{Ringdowns for black holes with scalar hair: the large mass case}

\author{Giovanni D'Addario}
\affiliation{School of Mathematical Sciences, University of Nottingham, Nottingham NG7 2RD, United Kingdom}
\affiliation{School of Physics and Astronomy, University of Nottingham, Nottingham NG7 2RD, United Kingdom}
\affiliation{Nottingham Centre of Gravity, Nottingham NG7 2RD, United Kingdom}
\author{Antonio Padilla}
\affiliation{School of Physics and Astronomy, University of Nottingham, Nottingham NG7 2RD, United Kingdom}
\affiliation{Nottingham Centre of Gravity, Nottingham NG7 2RD, United Kingdom}
\author{Paul M. Saffin}
\affiliation{School of Physics and Astronomy, University of Nottingham, Nottingham NG7 2RD, United Kingdom}
\affiliation{Nottingham Centre of Gravity, Nottingham NG7 2RD, United Kingdom}
\author{Thomas P. Sotiriou}
\affiliation{School of Mathematical Sciences, University of Nottingham, Nottingham NG7 2RD, United Kingdom}
\affiliation{School of Physics and Astronomy, University of Nottingham, Nottingham NG7 2RD, United Kingdom}
\affiliation{Nottingham Centre of Gravity, Nottingham NG7 2RD, United Kingdom}
\author{Andrew Spiers}
\affiliation{School of Mathematical Sciences, University of Nottingham, Nottingham NG7 2RD, United Kingdom}
\affiliation{Nottingham Centre of Gravity, Nottingham NG7 2RD, United Kingdom}

\date{\today}

\begin{abstract}
Deviations from general relativity can alter the quasi-normal mode (QNM) ringdown of perturbed black holes. It is known that a shift-symmetric (hence massless) scalar can only introduce black hole hair if it couples to the Gauss-Bonnet invariant, in which case the scalar charge is fixed with respect to the black hole mass and controlled by the strength of that coupling. The charge per unit mass decreases with the mass and can, therefore, be used as a perturbative parameter for black holes that are sufficiently large with respect to the scale suppressing the deviation from general relativity or the Standard model. We construct an effective field theory scheme for QNMs using this perturbative parameter to capture deviations from Kerr for both the background and the perturbations. We demonstrate that up to second order in the charge per unit mass, QNMs can be calculated by solving standard linearized perturbation equations for the Kerr metric with sources depending on solutions of the same equations up to first order. It follows that corrections to the QNM frequencies are heavily suppressed for sufficiently massive black holes, meaning that LISA is very unlikely to detect any evidence of scalar hair in ringdown signals.
\end{abstract}

\maketitle

\section{\label{sec:Introduction}Introduction}

The detection of gravitational waves (GWs) emitted by compact binary sources with the LIGO, Virgo, and KAGRA (LVK) interferometers \cite{LIGOScientific:2018mvr,LIGOScientific:2020ibl,LIGOScientific:2021usb,LIGOScientific:2021djp} has opened a new window on the physics of black holes and neutron stars. The data from the 90 binary mergers observed thus far have shed new light on the dynamical, strong field regime of gravity, leading to tests of general relativity (GR) \cite{LIGOScientific:2016lio,LIGOScientific:2018dkp,LIGOScientific:2019fpa,LIGOScientific:2020tif,LIGOScientific:2021sio, abbott2023population} that complement previous tests in the weak regime \cite{Will:2014kxa}, as well as tests performed using pulsars \cite{Stairs:2003eg,Wex:2014nva}. The extension of the ground-based detector network to include the Einstein Telescope \cite{Punturo:2010zz} and Cosmic Explorer \cite{Reitze:2019iox}, as well as the planned launch of the space-based detector LISA \cite{LISA:2017pwj}, promises to bring higher precision constraints on deviations from GR \cite{Maggiore:2019uih,Perkins:2020tra,Barausse:2020rsu,Evans:2021gyd,LISA:2022kgy}.

A key prediction of GR is encoded in uniqueness theorems that state that black holes belong to the Kerr family \cite{Kerr:1963ud} and are characterized exclusively by their mass and angular momentum \cite{Carter1971,Robinson1975}, neglecting electric charges. 
The detection of a black hole endowed with hair (scalar or of another kind) would, therefore, challenge 
GR or reveal the existence of new fundamental fields. Such a detection could involve the GWs emitted by a perturbed black hole as it reaches equilibrium, for example, in the ``ringdown" phase of a binary black hole merger. 

The ringdown emission of a black hole in GR is modeled using quasinormal modes (QNMs) (see \cite{Kokkotas:1999bd,Berti:2009kk,Konoplya:2011qq} for reviews), characterized by complex frequencies. That is, the time dependence of QNM perturbations $h_{\mu\nu}$ can be expressed as 
\begin{equation}
    h_{\mu\nu}\sim e^{-i\omega t}
\end{equation}
where $\omega=\omega_\mathcal{R} + i \omega_\mathcal{I}$. QNMs decay as they radiate into the black hole horizon and future null infinity; therefore, $\omega_\mathcal{I}$ is negative.
The QNM frequencies, $\omega$, depend only on the mass and angular momentum of the black hole, due to the uniqueness theorems. In principle, GW observations can be used to perform `black hole spectroscopy' \cite{Dreyer:2003bv,berti2006gravitational}, i.e., to determine the parameters of the black hole starting from the spectrum of the QNM frequencies; two or more frequencies would need to be extracted from GW data for this process to work. Thus far, confident detections of fundamental modes have been reported \cite{LIGOScientific:2016lio,LIGOScientific:2020tif,LIGOScientific:2021sio}. At the same time, there are ongoing debates regarding the significance of including higher overtones in ringdown models \cite{Giesler:2019uxc,Bhagwat:2019dtm,Baibhav:2023clw,Nee:2023osy} and the putative detections of overtones in LVK data \cite{Isi:2019aib,Cotesta:2022pci,Isi:2022mhy,Finch:2022ynt}. Similarly, there is interest in including second-order QNMs in models and the prospect of detecting the so-called quadratic second-order modes~\cite{ioka2007second, nakano2007second, cheung2023nonlinear, mitman2023nonlinearities, Lagos:2022otp,Redondo-Yuste:2023seq}. Future LVK observing runs, as well as the new detectors, will bring more precise measurements and are therefore expected to clarify the significance of these subtle GR effects. 

Advances in black hole spectroscopy could also provide a method of probing any additional black hole hair; indeed, the spectra of hairy black holes will generically differ from that of Kerr \cite{Franchini:2023eda}. Using a Bayesian analysis, Ref.~\cite{toubiana2023measuring} found that ringdown signals detected by LISA could be used to constrain deviations to GR within 10\% and, at best 1\%. 
A key question is: how large are the deviations one expects to get from well-motivated theories beyond GR and the Standard Model (SM)? 

Light scalar fields are ubiquitous in extensions of GR and the SM, so efforts to determine the QNM spectrum for black holes with scalar hair are receiving considerable attention. 
We highlight a few approaches that have recently been proposed to perform this task. The first consists of developing an effective field theory (EFT) for perturbations on a background spacetime before studying the dynamics of the perturbations in this model-independent framework. In one such study \cite{Franciolini:2018uyq}, the EFT for perturbations of static, spherically symmetric black holes with a spacelike scalar profile in scalar-tensor theories was formulated. After defining the metric perturbations in the traditional form of Regge and Wheeler \cite{ReggeWheeler1957}, and separating even- and odd-parity perturbations, the equations of motion for the perturbations can be obtained from the EFT, paving the way for calculations of QNM frequencies. In principle, the analysis can also proceed in the opposite direction, with direct measurements of QNM frequencies leading to constraints on the EFT parameters. This framework was subsequently extended to slowly rotating black holes \cite{Hui:2021cpm}. 

In a similar vein, an EFT for black hole perturbations in scalar-tensor theories about an arbitrary background geometry, with a timelike scalar profile, has been developed \cite{Mukohyama:2022enj}, and later applied to static, spherically symmetric backgrounds to derive the analogue of the Regge-Wheeler equation for the odd perturbations \cite{Mukohyama:2022skk}. From this equation, the QNM frequencies of the odd perturbations have been found for the stealth-Schwarzschild \cite{Mukohyama:2023xyf} and Hayward \cite{Konoplya:2023ppx} solutions. A further EFT-based approach has been developed in the context of higher-derivative models of gravity, followed by the derivation of a Teukolsky equation and the calculation of QNM frequencies for rotating black holes \cite{Cano:2019ore,Cano:2021myl,Cano:2023tmv,Cano:2023jbk}. The derivation of a Teukolsky equation, which is satisfied by the perturbed Weyl scalars (in terms of which gravitational perturbations can be expressed), represents a second approach to calculating QNM spectra in modified theories of gravity. A different approach to obtaining a modified Teukolsky equation has been derived for perturbations to non-Ricci-flat Petrov type I black hole backgrounds \cite{Li:2022pcy}. Finally, in \cite{Hussain:2022ins}, a perturbative treatment in the couplings controlling the deviations from GR has been used to develop a formalism for calculating QNM spectra with a Teukolsky equation.

In practice, `no-hair' theorems significantly limit the range of theoretical scenarios that lead to black hole hair. For example, a broad range of couplings of scalar fields to gravity are covered by no-hair theorems \cite{Hawking:1972qk,Bekenstein:1995un,Sotiriou:2011dz,Herdeiro:2015waa,Sotiriou:2015pka} for asymptotically flat, stationary black holes. For shift-symmetric scalars, there exists a no-hair theorem \cite{Hui:2012qt} for static and spherically symmetric black holes, which has been extended to slowly rotating black holes \cite{Sotiriou:2013qea}. It has been shown that the only coupling between a shift-symmetric, and hence massless, scalar and gravity that can evade this no-hair theorem, without leading to more degrees of freedom and without compromising local Lorentz symmetry \cite{Saravani:2019xwx}, is a linear coupling with the Gauss-Bonnet invariant, $\mathcal{G} = R^{\mu\nu\lambda\kappa}R_{\mu\nu\lambda\kappa} - 4R^{\mu\nu}R_{\mu\nu} + R^2$ \cite{Sotiriou:2013qea}. Even when one includes other shift-symmetric couplings, as we will discuss in more detail below, the scalar charge per unit mass $q$ is given by 
\begin{equation}
4 \pi q= \frac{\alpha}{M^2} \int_{\mathcal{H}} n_\mu \mathcal{G}^\mu \; ,
\label{eq:SSHorndeski_Dimensionless_Scalar_Charge}
\end{equation}
where $\alpha$ is the coupling controlling the $\phi {\cal G}$ interaction, $M$ is the black hole mass, $n^\mu$ is the normal to the horizon ${\cal H}$, and ${\cal G}^\mu$ is implicitly defined by ${\cal G}\equiv \nabla_\mu {\cal G}^\mu$ (recalling that ${\cal G}$ is a total divergence) \cite{Saravani:2019xwx}. \Cref{eq:SSHorndeski_Dimensionless_Scalar_Charge} together with the assumption that the new shift-symmetric terms do not introduce new length scales that are significantly larger than the one associated with $\alpha$, imply that the charge per unit mass scales as $M^{-2}$ for any given theory. Hence, large enough black holes, with respect to the scale associated with $\alpha$, will be very weakly charged.

Motivated by this observation, we develop an alternative approach for calculating deviations from the GR spectrum of QNM frequencies in the small charge/large mass limit.
Restricting attention to a metric and a massless scalar, we work with the shift-symmetric Horndeski action. This pragmatic approach is much simpler than theory agnostic methods, where one attempts to construct an EFT for black hole perturbations using symmetries, analogous to the EFT of inflation, even though the symmetries are far more limited in the black hole case. Of course, as the most general action that yields second-order field equations (up to field redefinitions), the shift-symmetric Horndeski action automatically includes all interactions for the chosen field content. 
%\comment{From an EFT perspective, one may add other terms to the Lagrangian density but, by restricting to Horndeski, we are guaranteed not to introduce any extra degrees of freedom. For example, including higher powers of curvature in the form of an $f(R)$ theory is equivalent to having an additional scalar along with the metric degree of freedom (ref).}. 
Making use of the scaling of the charge with the black hole mass, we then perform a double expansion in the charge (per unit mass) and in linear dynamical perturbations. This reduces the action to a well-defined EFT, faithfully describing perturbations in the small charge/large mass limit, and allows us to define a hierarchical set of perturbative field equations.

We expect our framework to be particularly well suited to LISA observations of supermassive (and intermediate-mass) black hole mergers~\cite{LISAConsortiumWaveformWorkingGroup:2023arg}. Observations of black holes in the range of a few solar masses by LVK and other strong field observations tend to yield constraints on the length scale of new couplings of the order of km \cite{Yagi:2012gp,Saffer:2021gak,Nair:2019iur, Perkins:2021mhb,Yamada:2019zrb,Wang:2021jfc,Lyu:2022gdr,Fernandes:2022kvg}. In particular, this would imply that supermassive black holes of $10^5-10^6$ solar masses would indeed have very small charges.

Similar considerations have recently been used to simplify the problem of modelling extreme mass ratio inspirals (EMRIs) in theories with an additional scalar field \cite{Maselli:2020zgv,Barsanti:2022ana,Barsanti:2022vvl,Spiers:2023cva}. The deviation due to any charge of the supermassive primary has been shown to be entirely negligible, and instead, observations are expected to place stringent constraints on the charge of the secondary \cite{Maselli:2020zgv}. This is in stark contrast with conventional thinking about EMRIs, which sees the motion of the secondary as a probe of the spacetime of the primary. In the case of QNMs with LISA, however, the focus is on a single, postmerger excited black hole. This is the case we are considering here to assess if LISA measurements of the QNM ringdown can be used to detect deviations from the Kerr spacetime and GR.

Throughout this paper, unless otherwise specified, we employ geometric units with $c = G = 1$.

\section{\label{sec:Theoretical_background}Theoretical background}

The action for shift-symmetric Horndeski gravity, invariant under a shift $\phi \to \phi + \mathrm{constant}$ of the scalar $\phi$, is
\begin{equation}
    S_H=\frac{1}{16 \pi} \sum_{i=2}^{5} \int d^{4} x \sqrt{-g} \mathcal{L}_{i}+S_{M} \;.
    \label{eq:SSHorndeski_Action}
\end{equation}
where the Lagrangians $\mathcal{L}_{i}$ are given by
\begin{align}
\mathcal{L}_{2}&=K(X) \notag\\
\mathcal{L}_{3}&=-G_{3}(X) \square \phi \notag\\
\mathcal{L}_{4}&=G_{4}(X) R+G_{4X}(X)\big[(\square \phi)^{2}-\big(\nabla_{\mu} \nabla_{\nu} \phi\big)^{2}\big] \notag\\
\mathcal{L}_{5}&=G_{5}(X) G_{\mu \nu} \nabla^{\mu} \nabla^{\nu} \phi  - \frac{G_{5X}}{6}\big[\big(\square \phi)^{3} \notag\\
& -3 \square \phi\big(\nabla_{\mu} \nabla_{\nu} \phi\big)^{2}+2\big(\nabla_{\mu} \nabla_{\nu} \phi\big)^{3}\big] \; ,
\label{eq:SSHorndeski_Lagrangians}
\end{align}
with $K$ and $G_i$ arbitrary functions of $X=-\frac{1}{2} \nabla_{\mu} \phi \nabla^{\mu} \phi$ in the shift-symmetric case \cite{Sotiriou:2014pfa}. Above, we have employed the notation $f_{X} = df/dX$, $(\nabla_{\mu} \nabla_{\nu} \phi)^{2} = \nabla_{\mu} \nabla_{\nu} \phi \nabla^{\mu} \nabla^{\nu} \phi$, $(\nabla_{\mu} \nabla_{\nu} \phi)^{3} = \nabla_{\mu} \nabla_{\nu} \phi \nabla^{\nu} \nabla^{\lambda} \phi \nabla_{\lambda} \nabla^{\mu} \phi$, while $R$ and $G_{\mu\nu}$ are the Ricci scalar and the Einstein tensor, respectively. The matter action is $S_{M}$; here, we do not consider additional matter beyond the scalar $\phi$.

Shift-symmetric theories described by \cref{eq:SSHorndeski_Action} have been divided into three classes, according to whether they admit solutions with the trivial scalar field profile $\phi=0$ in flat space or curved spacetimes \cite{Saravani:2019xwx}. The first class consists of theories allowing $\phi=0$ solutions and, therefore, all GR solutions. The second class comprises theories that admit Minkowski as a solution with $\phi=0$ but otherwise have solutions with nontrivial scalar field configurations. The third class does not admit flat space with constant scalar as a solution and is in clash with Local Lorentz symmetry. The first two classes are actually related; if $\Tilde{\mathcal{L}}$ belongs to the first class and $\mathcal{L}$ is in the second class, then
\begin{equation}
    \mathcal{L} = \Tilde{\mathcal{L}} + \alpha \phi \mathcal{G} \;,
    \label{eq:SSHorndeski_Lagrangian_Class2}
\end{equation}
where $\alpha$ is a coupling constant with dimensions length squared. Moreover, the Lagrangian $\Tilde{\mathcal{L}}$ can be written as the sum of the terms in \cref{eq:SSHorndeski_Lagrangians}, with coefficients $\Tilde{G}_i(X)$ satisfying specific analyticity conditions as $X \to 0$ \cite{Saravani:2019xwx}. These conditions are met for the choice 
\begin{align}
    \tilde{K}(X) & = X + \mathcal{O}(X^2) ,\notag\\
    \tilde{G}_3(X) & = \tau_3 X + \mathcal{O}(X^2),\notag\\
    \tilde{G}_4(X) & = 1 + \tau_4 X + \mathcal{O}(X^2),\notag\\
    \tilde{G}_5(X) & = \tau_5 X +  \mathcal{O}(X^2) \; .
    \label{eq:SSHorndeski_Coefficients}
\end{align}
The coupling constant preceding the leading order term in $\tilde{K}(X)$ can be set to one without loss of generality by rescaling the scalar. $\tilde{K}$ does not include a constant term, as that corresponds to a cosmological constant which we assume to vanish; constant terms in $\tilde{G}_3(X)$ and $\tilde{G}_5(X)$ give total derivatives, so their values can be safely set to zero, and the expansions start at order $X$; $\tilde{G}_4$ starts with a constant term whose value is chosen to be one, so that we get GR as $X\to0$. 

We can then write the Lagrangian for our shift-symmetric theory as in \cref{eq:SSHorndeski_Lagrangian_Class2}, 
where $\Tilde{\mathcal{L}}$ admits GR solutions, and the full theory admits curved spacetimes with nontrivial scalar profiles. This statement highlights the importance of the linear coupling of the scalar with the Gauss-Bonnet invariant, which, as discussed in \cref{sec:Introduction}, is the unique term in Lorentz invariant, shift-symmetric Horndeski theories which allows for the evasion of a no-hair theorem, leading to solutions with nontrivial scalar field profiles. We note that the action written in this form is still shift-symmetric, as $\mathcal{G} =  \nabla_\mu \mathcal{G}^\mu$ is a total divergence in four dimensions.

Bounds on the value of $\alpha$ have been obtained from astrophysical observations of black hole low mass x-ray binaries \cite{Yagi:2012gp} and neutron stars \cite{Saffer:2021gak}, and from gravitational wave observations of binary black holes \cite{Nair:2019iur, Perkins:2021mhb, Yamada:2019zrb, Wang:2021jfc} and neutron star black hole binaries \cite{Lyu:2022gdr} in the first three LVK observing runs. The strongest bound to date has been found by studying the minimum mass of a black hole in theories with a nonminimal coupling between $\phi$ and $\mathcal{G}$; for the linear coupling $\bar\alpha \phi\mathcal{G}$, with $\bar\alpha =\frac{1}{16 \pi}\alpha $, the upper bound is $\sqrt{\bar\alpha} \lesssim (0.82 \pm 0.03)$ km \cite{Fernandes:2022kvg}.

The coupling constants $\alpha$, $\tau_3$ and $\tau_4$ have dimensions of length squared, while $\tau_5$ has dimensions of length to the power of 4. These couplings are expected to be related in a natural theory. Indeed, the Lagrangian for an EFT with natural couplings can be written schematically as 
$\mathcal{L}=\sum_k \frac{f^4}{\mu^k} \mathcal{O}_k[\frac{\varphi}{\nu}]$ where $f, \mu, \nu$ are the corresponding mass scales of the theory, $\varphi$ describes the canonically normalized bosonic degrees of freedom, and the operators $ \mathcal{O}_k$ are assumed to contain $k$ derivatives and coefficients of order one \cite{Burgess:2007pt}. If we focus on the structure of the canonically normalized kinetic terms, we can easily identify $f \sim \sqrt{\mu \nu}$ \cite{Gavela:2016bzc}. All of this suggests that natural Horndeski theories can be written in the form $\mathcal{L}=\mu^2 M_{P}^2 \mathcal{F}\left(g_{\alpha\beta}, \frac{R_{\alpha\beta\gamma\delta}}{\mu^2}, \phi, \frac{\nabla_\alpha \phi}{\mu},  \frac{\nabla_\alpha \nabla_\beta \phi}{\mu^2}\right)$ where the scalar $\phi$ is taken to be dimensionless and $\mathcal{F}$ admitting an expansion in $ \frac{R_{\alpha\beta\gamma\delta}}{\mu^2}, \phi, \frac{\nabla_\alpha \phi}{\mu}$ and $ \frac{\nabla_\alpha \nabla_\beta \phi}{\mu^2}$ with order one coefficients. Note that we have also temporarily reinstated the Planck scale, $M_P^2 =1/8 \pi G$. When applied to the Lagrangian of \cref{eq:SSHorndeski_Lagrangian_Class2}, naturalness considerations now imply that $\alpha\sim \tau_3 \sim \tau_4 \sim 1/\mu^2$ and $\tau_5 \sim 1/\mu^4$. However, as we will see later, our effective description remains under control even when the $\tau_i$ are much larger than their natural values. 

In a standard EFT approach one could consistently consider higher order operators that yield higher order equations of motion. Such operators should remain as small corrections to the leading order results presented. This is expected to be the case, since on grounds of naturalness, each new derivative comes with a power of $1/\mu\sim \sqrt{\alpha} \sim \sqrt{q}$. If such terms do become dominant, we have excited the new  degrees of freedom associated with those higher derivatives, representing a breakdown of our effective description in terms of the degrees of freedom of a massless scalar and a massless tensor.

In the absence of any additional matter content in the theory beyond the scalar $\phi$, the field equation for the metric is of the form
\begin{equation}
    G_{\mu\nu} = T^\phi_{\mu\nu} \; ,
    \label{eq:SSHorndeski_Metric_EOM}
\end{equation}
where $T^\phi_{\mu\nu}$ is the energy-momentum tensor for $\phi$. The full equation for the general Horndeski theory without shift-symmetry can be found in \cite{Kobayashi:2011nu}; the contribution of the $\phi \mathcal{G}$ term to \cref{eq:SSHorndeski_Metric_EOM} is given in \cite{Sotiriou:2014pfa}.

The equation of motion for the scalar in the theory $\Tilde{\mathcal{L}}$ belonging to the first class of shift-symmetric Horndeski theories can be written as the conservation law
\begin{equation}
    \nabla_\mu \Tilde{J}^\mu = 0 \;,
\end{equation}
with the current $\Tilde{J}^\mu$ found in \cite{Sotiriou:2014pfa}.
The scalar equation for \cref{eq:SSHorndeski_Lagrangian_Class2} is then
\begin{equation}
    \nabla_\mu( \Tilde{J}^\mu - \alpha\mathcal{G}^\mu)  = 0 \;.
    \label{eq:SSHorndeski_Scalar_EOM}
\end{equation}

It has been shown that for stationary black hole solutions of mass $M$ belonging to the two subclasses of shift-symmetric Horndeski theories described above, the dimensionless scalar charge $q$ obeys \cref{eq:SSHorndeski_Dimensionless_Scalar_Charge}. As $\alpha \to 0$, the scalar charge vanishes, and the theory \cref{eq:SSHorndeski_Lagrangian_Class2} reduces to a shift-symmetric Horndeski belonging to the first class, resulting in GR solutions with constant $\phi$. Therefore, the beyond-GR solutions with nontrivial scalar field profiles are continuously connected to the GR ones, in the limit that $\alpha$, and therefore $q$, go to zero. Consequently, beyond-GR behavior in this class of shift-symmetric theories is inversely proportional to the square of the black hole mass. These statements motivate the development of a perturbative treatment for the ringdown of black holes with scalar hair about a perturbed GR solution, with $q$ serving as the perturbative parameter.
Note that our formalism does not capture solutions that are parametrically disconnected from those in GR.

\section{\label{sec:Perturbative_Framework}Perturbative framework}

The analysis of QNMs in shift-symmetric Horndeski theories arises naturally in a perturbative formalism, allowing us to calculate the deviations to the GR QNMs in terms of $q$, or $\alpha$. Our approach consists of a double expansion; initially, we consider a dynamical perturbation, parameterized by $\varepsilon$, of a GR background. 
Following the standard first-order procedure, our formalism results in the GR description of black hole ringdown, with an additional scalar field perturbation. We introduce nondynamical perturbations in $q$ in the resulting system, generated by the non-GR terms in the action \cref{eq:SSHorndeski_Action}. We employ the superscripts $(n, m)$ to denote the order of a quantity with respect to $(\varepsilon, q)$.

We limit our expansion to linear perturbations in $\varepsilon$, as is usual for QNMs, while we go up to quadratic order in $q$. The perturbed scalar and metric are then
\begin{align}
    \phi & =  q \phi^{(0, 1)} +  \varepsilon \phi^{(1, 0)} +  q^2\phi^{(0, 2)} + \varepsilon q  \phi^{(1, 1)} \label{eq:SSHorndeski_SF_Perturbed}\\
    g_{\mu\nu} & = \bar{g}_{\mu\nu} + \varepsilon h_{\mu\nu}^{(1,0)}  + q^2 h^{(0, 2)}_{\mu\nu} + \varepsilon q h^{(1,1)}_{\mu\nu} \; ,
\label{eq:SSHorndeski_Metric_Perturbed}
\end{align}
where $\bar g_{\mu\nu}$ is the Kerr metric and we set the background scalar field $\bar \phi$ to zero using shift-symmetry. The perturbation $qh^{(0, 1)}_{\mu\nu}$ is taken to be zero as it is not a physically excited mode; indeed, in the limit $\varepsilon \to 0$, corrections in $q$ to the metric are due to the scalar perturbations and their backreaction, and therefore appear at order $q^2$, as is well known from studying static configurations \cite{Sotiriou:2014pfa}. Including higher order corrections in $q$ is straightforward but unnecessary considering how small $q$ is expected to be for large black holes. The size of $\varepsilon$ is linked to the amplitude of the perturbations. It is expected to be larger than $q$ at the onset of the ringdown, but as QNMs decay the stationary correction 
 related to $q$ will dominate.

 \section{Perturbation equations}

 By perturbing \cref{eq:SSHorndeski_Metric_EOM} and \cref{eq:SSHorndeski_Scalar_EOM} with respect to the background metric and scalar field, using \cref{eq:SSHorndeski_Metric_Perturbed,eq:SSHorndeski_SF_Perturbed}, we obtain the following equations:
\begin{align}
  \delta G_{\mu\nu}&[h^{(1, 0)}_{\mu\nu}]     = 0\; , \label{eq:Perturbed_Metric_EOM_Order_10}\\
  \delta G_{\mu\nu}&[h^{(0, 2)}_{\mu\nu}]   =
 - \frac{1}{2}\alpha^{(0, 1)} \big[\bar g_{\rho \mu } \bar g_{\delta\nu} + \bar g_{\rho\nu} \bar g_{\delta\mu} \big] 
 \notag\\
& \cdot \bar\nabla_\sigma \big(\bar\nabla_\gamma \phi^{(0,1)} \bar\epsilon^{\lambda\eta\rho\sigma} \bar\epsilon^{ \alpha \beta \gamma\delta} \bar R_{\lambda \eta\alpha \beta}\big) \notag\\
 & + \frac{1}{2}\bar\nabla_{\mu} \phi^{(0, 1)} \bar\nabla_{\nu} \phi^{(0, 1)} - \frac{1}{4} \big(\bar\nabla_{\alpha} \phi^{(0,1)}\big)^2\bar g_{\mu \nu} \notag\\
&
+ \tau_4 \big\{\bar\Box\phi^{(0,1)}\bar\nabla_\mu \bar\nabla_\nu \phi^{(0,1)} - \bar\nabla_\lambda \bar\nabla_\mu \phi^{(0,1)} \bar\nabla^\lambda \bar\nabla_\nu \phi^{(0,1)} \notag \\
& -\frac{1}{2}\bar g_{\mu\nu}\big[(\bar\Box\phi^{(0,1)})^2 - (\bar\nabla_\alpha \bar\nabla_\beta\phi^{(0,1)})^2    \big] \notag \\
& - \bar R_{\mu \alpha \nu \beta} \bar\nabla^\alpha \phi^{(0,1)} \bar \nabla^\beta \phi^{(0,1)} \big\}\; ,
\label{eq:Perturbed_Metric_EOM_Order_02}\\
\delta G_{\mu\nu}&[h^{(1, 1)}_{\mu\nu}]  = 
- \frac{1}{2}\alpha^{(0, 1)} \big[\bar g_{\rho \mu } \bar g_{\delta\nu} + \bar g_{\rho\nu} \bar g_{\delta\mu}\big] \notag\\
& \cdot  \bar\nabla_\sigma \big(
  \bar\nabla_\gamma \phi^{(1,0)} \bar\epsilon^{\lambda\eta\rho\sigma} \bar\epsilon^{ \alpha \beta \gamma\delta} \bar R_{\lambda \eta\alpha \beta} \big) \notag\\
& 
+ \bar\nabla_{(\mu} \phi^{(0, 1)} \bar\nabla_{\nu)} \phi^{(1, 0)} - \frac{1}{2} \bar\nabla_{\alpha} \phi^{(0,1)}\bar\nabla^{\alpha} \phi^{(1,0)} \bar g_{\mu \nu} \notag \\
&
+ \tau_4 \big\{\bar\Box\phi^{(0,1)}\bar\nabla_\mu \bar\nabla_\nu \phi^{(1,0)} + \bar\Box\phi^{(1,0)}\bar\nabla_\mu \bar\nabla_\nu \phi^{(0,1)} \notag\\
& - 2\bar\nabla_\lambda \bar\nabla_{(\mu} \phi^{(1,0)} \bar\nabla^\lambda \bar\nabla_{\nu)} \phi^{(0,1)} \notag \\
& -\bar g_{\mu\nu}\big[\bar\Box\phi^{(0,1)}\bar\Box\phi^{(1,0)} - \bar\nabla_\alpha \bar\nabla_\beta\phi^{(0,1)}\bar\nabla^\alpha \bar\nabla^\beta\phi^{(1,0)}  \big] \notag \\
& - 2\bar R_{\mu \alpha \nu \beta} \bar\nabla^{(\alpha} \phi^{(1,0)} \bar \nabla^{\beta)} \phi^{(0,1)} \big\}
\; .
 \label{eq:Perturbed_Metric_EOM_Order_11}\\
 \bar\Box \phi^{(0, 1)} &  =  -\alpha^{(0, 1)} \bar R_{\mu\nu\rho \sigma} \bar R^{ \mu \nu \rho\sigma} \, ,
\label{eq:Perturbed_SF_EOM_Order_01} \\
\bar\Box \phi^{(1, 0)} & = 0\,. \label{eq:Perturbed_SF_EOM_Order_10}  
\end{align}
$\bar\nabla_\mu$ and $\bar R_{\mu \alpha \nu \beta}$ are associated with the background metric $\bar g_{\mu\nu}$, while we introduce $\alpha^{(0,1)}$ by $\alpha=q\alpha^{(0,1)}$, to make it easier to keep track of perturbation orders, given that $\alpha\sim\mathcal{O}(q)$ by \cref{eq:SSHorndeski_Dimensionless_Scalar_Charge}. The linearized Einstein operator acting on the metric perturbations in \cref{eq:Perturbed_Metric_EOM_Order_10,eq:Perturbed_Metric_EOM_Order_02,eq:Perturbed_Metric_EOM_Order_11} is 
\begin{equation}
    \delta G_{\mu\nu} = \delta \bigg[R_{\mu\nu}- \frac{1}{2}g_{\mu\nu}R \bigg] = \bigg(\tensor{\bar g}{_\mu^\alpha} \tensor{\bar g}{_\nu^\beta} - \frac{1}{2} \bar g_{\mu\nu} \bar g^{\alpha\beta}\bigg)\delta R_{\alpha\beta} \; , 
\end{equation}
with the Lichnerowicz operator
\begin{equation}
    \delta R_{\mu\nu}[h] = -\frac{1}{2}\Big(\bar\Box\tensor{h}{_\mu_\nu} - 2\tensor{h}{^\alpha_{(\mu;\nu)}_\alpha}  + \tensor{h}{^\alpha_\alpha_{;\mu \nu}} \Big)
\end{equation}
acting on a general perturbation $h_{\mu\nu}$. 

As can be seen from \cref{eq:Perturbed_Metric_EOM_Order_10}, $h^{(1,0)}_{\mu\nu}$ is a solution of the homogeneous version of \cref{eq:Perturbed_Metric_EOM_Order_11,eq:Perturbed_Metric_EOM_Order_02} and hence one only needs to solve for $h^{(1,1)}_{\mu\nu}$ to determine what the detector would measure. The source in \cref{eq:Perturbed_Metric_EOM_Order_11} is therefore nonstationary, so the particular solution $h^{(1,1)}_{\mu\nu}$ will contain detectable non-GR QNM perturbations. Moreover, 
\cref{eq:Perturbed_Metric_EOM_Order_02,eq:Perturbed_Metric_EOM_Order_11} are sourced by $\phi^{(0, 1)}$ and $\phi^{(1, 0)}$ but not by $\phi^{(0,2)}$ and $\phi^{(1,1)}$, which is why we have omitted the equation for the latter two. We assume $\phi^{(1,0)}$ contains all the homogeneous dynamical QNM content, being a solution to \cref{eq:Perturbed_SF_EOM_Order_10}, and $\phi^{(0,1)}$ is the particular solution to \cref{eq:Perturbed_SF_EOM_Order_01}, which is stationary. The particular solution to $h^{(0,2)}$ in \cref{eq:Perturbed_Metric_EOM_Order_02} is entirely stationary and does also not contribute to $h^{(1,1)}_{\mu\nu}$ at the orders we consider; therefore, its contribution is not detectable, and we do not need to calculate it.

We now focus on the source terms in \cref{eq:Perturbed_Metric_EOM_Order_11} and their magnitude. These include a $\tau_4$ contribution but no contributions from $\tau_3$ and $\tau_5$. It is important that the $\tau_i$ are dimensionful and that, in obtaining the equations above we have assumed they can each be as large as order one in units of the astrophysical scale $M$. Even for such large couplings, the $\tau_3$ and $\tau_5$ contributions to \cref{eq:Perturbed_Metric_EOM_Order_11} are already beyond $\varepsilon q$. This is because these terms contain at least two copies of the scalar field. However, we have constructed our expansion under the (small charge) assumption $\alpha \sim q M^2$. Assuming that the scale associated with $\alpha$ is roughly the same as those associated with the $\tau_i$, would imply $\tau_3\sim\tau_4\sim q M^2$ and $\tau_5 \sim q^2 M^4$. The absence of the scale hierarchy is indeed what one expects from the naturalness arguments we presented earlier, and it introduces a further suppression of the $\tau_i$ contributions. In particular, it pushes the $\tau_4$ contribution in \cref{eq:Perturbed_Metric_EOM_Order_11} to order $\varepsilon q^2$. One is then left with the much simpler system
\begin{align}
\label{eq:Perturbed_Metric_EOM_Order_11_new}
\delta G_{\mu\nu}&[h^{(1, 1)}_{\mu\nu}]  = 
- \frac{1}{2}\alpha^{(0, 1)} \big[\bar g_{\rho \mu } \bar g_{\delta\nu} + \bar g_{\rho\nu} \bar g_{\delta\mu}\big] \\
& \cdot  \bar\nabla_\sigma \big(
  \bar\nabla_\gamma \phi^{(1,0)} \bar\epsilon^{\lambda\eta\rho\sigma} \bar\epsilon^{ \alpha \beta \gamma\delta} \bar R_{\lambda \eta\alpha \beta} \big) \notag\\
& 
+ \bar\nabla_{(\mu} \phi^{(0, 1)} \bar\nabla_{\nu)} \phi^{(1, 0)} - \frac{1}{2} \bar\nabla_{\alpha} \phi^{(0,1)}\bar\nabla^{\alpha} \phi^{(1,0)} \bar g_{\mu \nu} \notag
\; ,
 \\
 \bar\Box \phi^{(0, 1)} &  =  -\alpha^{(0, 1)} \bar R_{\mu\nu\rho \sigma} \bar R^{ \mu \nu \rho\sigma} \, ,
 \label{eq:Perturbed_SF_EOM_Order_01_2} \\
 \bar\Box \phi^{(1, 0)} & = 0\,.  
 \label{eq:Perturbed_SF_EOM_Order_10_2} 
\end{align}
Remarkably, the only remaining contributions come from the $\alpha \phi \mathcal{G}$ term in the action.  

Note, $\delta G_{\mu\nu}$ is computed in a background Kerr spacetime and therefore has the important property that it admits a separable Teukolsky equation~\cite{Teukolsky:1973ha, wald1973perturbations}, even beyond first-order~\cite{Spiers-Second-orderTeuk, green2020teukolsky}. This property was used by Ref.~\cite{Hussain:2022ins} to derive a separable Teukolsky equation in theories beyond GR. Similarly, one can extract sourced, separable Teukolsky equations from \cref{eq:Perturbed_Metric_EOM_Order_10,eq:Perturbed_Metric_EOM_Order_02,eq:Perturbed_Metric_EOM_Order_11}.

Indeed, from \cref{eq:Perturbed_Metric_EOM_Order_11_new} (or similarly \cref{eq:Perturbed_Metric_EOM_Order_11} if the $\tau_4$ contributions were retained) we can straightforwardly derive the frequency of the solutions $h_{\mu\nu}^{(1,1)}$ from the time dependency of the source terms. All parts of the source are stationary, apart from $\phi^{(1,0)}$. The solutions $\phi^{(1,0)}$ are $s=0$ QNMs. Hence the frequency of $h_{\mu\nu}^{(1,1)}$ is also the frequency of $s=0$ QNMs. Now we examine how long-lived the $s=0$ QNMs are compared to the $s=2$ QNMs we expect in GR. Looking at the dominant $\ell=2$, $m=2$, $n=0$ modes in Schwarzschild~\cite{Berti:2009kk, BHPToolkit}: for $s=0$, $\omega=0.483644 - 0.0967588 i$; whereas for $s=2$, $\omega = 0.373672 - 0.0889623 i$. That is, the $s=2$ mode decays slower as the imaginary part of its frequency is less negative, but only by around $10\%$. In comparison, the overtone $s=2$, $\ell=2$, $m=2$, $n=1$ has a frequency of $\omega=0.346711 - 0.273915 i$, and so decays much more rapidly. Similarly, the imaginary parts of the quadratic QNMs are double that of the linear QNM and, hence, also decay more rapidly. Therefore, the non-GR QNMs generated in our formalism are longer lived than any other contribution apart from the fundamental linear QNMs. If the non-GR QNMs were not suppressed by $q$ being small, then it would be the first detectable effect after the fundamental frequency. While we have used the Schwarzschild limit as an example here, the relationship between the frequency of the fundamental mode for $s=0$ and $s=2$, the overtones, and the quadratic QNMs, is similar for nonzero $a$.

It is clear that a nontrivial $\phi^{(0, 1)}$ is generated from the source in \cref{eq:Perturbed_SF_EOM_Order_01_2}; this is the usual scalar profile of a scalar charged black hole due to the scalar-Gauss-Bonnet coupling in the action. However, the sources vanish if $\phi^{(1, 0)}=0$. This implies that the magnitude of the deviations from GR depends crucially on the amplitude of the scalar perturbation at the onset of the ringdown. This could introduce a further suppression if the merger does not lead to large amplitudes for $\phi^{(1, 0)}$. It is worth stressing that, during the inspiral, scalar radiation is controlled by the difference in the scalar charges of the two initial black holes. The scalar field of these black holes will be $\mathcal{O}(q)$ in our setup. Hence, the reasonable assumption there is to set $\phi^{(1, 0)}=0$ \cite{Witek:2018dmd}. It is tempting to do the same here, invoking the argument that, in the $\alpha \to 0$ limit, one recovers GR with a minimally coupled scalar field, and the latter would not get excited in a merger. In that case, there would be no correction at $\mathcal{O}(\varepsilon q)$ to the tensor modes with respect to GR in our approximation. We will remain more conservative and not set $\phi^{(1, 0)}=0$ to remain agnostic about the merger and the prospect that nonlinearities might lead to a very large initial scalar amplitude.

Finally, we consider the question of gauge dependency. An infinitesimal diffeomorphism generated by the vector field $\chi^a = q \chi^{(0,1)a} + \varepsilon \chi^{(1,0)a} + q^2 \chi^{(0,2)a} + \varepsilon q \chi^{(1,1)a}$ shifts the scalar by
$\phi \to \phi + \delta \phi = \phi +  \mathsterling_\chi \phi$, where $\mathsterling$ is a Lie derivative. For the perturbed scalar field \cref{eq:SSHorndeski_SF_Perturbed}, the diffeomorphism leads to
\begin{align}
    \delta \phi & = \delta \bar\phi + q\delta \phi^{(0, 1)}+ \varepsilon\delta\phi^{(1,0)} + \mathcal{O}(q^2, \varepsilon q, \varepsilon^2) \notag\\
    & = q \mathsterling_{\chi^{(0,1)}}\bar\phi + \varepsilon  \mathsterling_{\chi^{(1,0)}} \bar\phi + \mathcal{O}(q^2, \varepsilon q, \varepsilon^2) \; .
\end{align}
At order $q$, the gauge dependence of the perturbed scalar is then $\delta \phi^{(0,1)} = \mathsterling_{\chi^{(0,1)}} \bar\phi=0$, and so the $\phi^{(0,1)}$ solution is gauge invariant. The same is true for the $\phi^{(1,0)}$, which has gauge dependence $\delta \phi^{(1,0)} = \mathsterling_{\chi^{(1,0)}} \bar\phi=0$. We note that the higher order perturbations $\phi^{(0,2)}$ and $\phi^{(1,1)}$, which we do not need to find, are not gauge invariant.

We also note that \cref{eq:Perturbed_Metric_EOM_Order_10}, by construction, is the same as in GR; hence the solutions $ h^{(1,0)}_{\mu\nu}$ will be GR solutions, in the form of QNMs. At this level, the gauge freedom is 
\begin{equation}
    h^{(1,0)}_{\mu\nu} \to h^{(1,0)}_{\mu\nu} + \mathsterling_{\chi^{(1,0)}} \bar g_{\mu\nu} \; ,
\end{equation}
as in linearized GR \cite{Pound_2021p}. As the source terms in \cref{eq:Perturbed_Metric_EOM_Order_02,eq:Perturbed_Metric_EOM_Order_11} act on the gauge invariant perturbations $\phi^{(0,1)}$ and $\phi^{(1,0)}$, we deduce that the gauge dependence of $h^{(0,2)}_{\mu\nu}$ and $h^{(1,1)}_{\mu\nu}$ is given by 
\begin{align}
    h^{(1,1)}_{\mu\nu} &\to h^{(1,1)}_{\mu\nu} +  \mathsterling_{\chi^{(0,1)}} h^{(1,0)}_{\mu\nu}+ \mathsterling_{\chi^{(1,1)}} \bar g_{\mu\nu}  \\
    h^{(0,2)}_{\mu\nu} &\to h^{(0,2)}_{\mu\nu} + \mathsterling_{\chi^{(0,2)}} \bar g_{\mu\nu} +\frac12\mathsterling_{\chi^{(0,1)}}^2 \bar g_{\mu\nu} \; ,
\end{align}
where we have used $h^{(0,1)}_{\mu\nu}=0$.

\section{\label{sec:Discussion}Discussion}

In this paper, we have developed a framework for investigating the impact of new gravitational physics on the quas-normal modes emitted during the ringdown phase of gravitational wave emission. The framework is particularly well suited to LISA, which is expected to detect gravitational wave emission from black holes with masses in the range of $10^3$ to $10^9$ solar masses. Owing to the high signal-to-noise ratio, potential LISA QNM measurements are often touted as an effective probe of new gravitational physics.

Our method makes use of the shift-symmetric Horndeski action, known to be the most general action yielding second order field equations for a metric and a massless scalar field (up to field redefinitions). Assuming that the solutions are continuously connected to those of GR, the field equations are then expanded order by order in two small parameters: the scalar charge per unit of black hole mass alongside the standard dynamical QNM perturbations of the metric and the scalar field. This gives rise to a well-defined EFT for black hole perturbations in the small charge/large mass limit.

Two immediate conclusions follow from our analysis: first, the linear coupling to the Gauss-Bonnet invariant is the only deviation from GR that persists in the small charge/large mass limit, provided that the length scales associated with the other leading order couplings are not {\it significantly} larger than that of the Gauss-Bonnet coupling; second, for small enough charge per unit mass, the metric perturbation equations reduce to those of GR sourced by the leading order fluctuations in the scalar field. 

The reference scale here is the characteristic length scale of the coupling to the Gauss-Bonnet invariant, which controls the scalar charge. This scale is already constrained to be of order km or smaller by past and current observations \cite{Yagi:2012gp,Saffer:2021gak,Nair:2019iur, Perkins:2021mhb, Yamada:2019zrb, Wang:2021jfc,Lyu:2022gdr,Fernandes:2022kvg}. Hence black holes with masses $M \gtrsim 10^3 M_\odot$ are expected to have negligible charges per unit mass, $q\lesssim 10^{-6}$, and their perturbations will be comfortably described within our formalism once the amplitude falls below the same scale. Corrections to the QNM frequencies will arise from source terms suppressed by $q$ times the amplitude of the scalar perturbation in the limit of zero charge.
This makes it very unlikely that LISA would be able to detect evidence of the charge in the ringdown of black holes.

Finally, one could consider applying our formalism to much smaller black holes, down to 100 solar masses or less. In this instance, $q$ will be much larger than for supermassive black holes, but still small enough to be a good expansion parameter. One could then calculate the shift in QNM frequencies, quantifying the deviations from Kerr to how they scale with the mass. If the precision with which QNMs can be measured by each detector is also factored in, we could investigate at which mass one expects QNMs to start giving useful constraints for deviations from Kerr stemming from scalar charge. We will address this in future work.

\section{Acknowledgements}

This work made use of the BHPToolkit. We acknowledge partial support from the STFC Consolidated Grants No. ST/V005596/1, No. ST/T000732/1, and No. ST/X000672/1.

\bibliography{references}

\end{document}